 \documentclass[aps,prl,twocolumn]{revtex4-1}

\usepackage{amsmath}
\usepackage{upgreek}
\usepackage{graphicx}% Include figure files
\usepackage{dcolumn}% Align table columns on decimal point
\usepackage{bm}% bold math
\usepackage{xcolor}

\usepackage{amssymb}
\usepackage{upgreek}

\usepackage{etoolbox}
\makeatletter
\def\bibsection{%
    \vspace{\baselineskip}%
    \global\let\bibsection@sw\@empty
}
\makeatother

\begin{document}

\preprint{APS/123-QED}
\title{Programmable branched flow of light}
%\title{Controlled channeling of anisotropic branched light flow in  \\ engineered liquid crystal landscapes}% 

\author{Shan-shan Chang$^{1,2}$}
\author{Daxing Xiong$^{3}$}
\author{Ze-huan Zheng$^{1}$}
\author{Li-Wei Wang$^{4,5}$}
\author{Yan-qing Lu$^{6}$}
\email{yqlu@nju.edu.cn}
\author{\\Lu-Jian Chen$^{7}$}
\email{lujianchen@xmu.edu.cn}
\author{Jian-Hua Jiang$^{4,5}$}
\email{jhjiang3@ustc.edu.cn}
\author{Jin-hui Chen$^{1,2}$}
\email{jimchen@xmu.edu.cn} 
%\author{\\Jian-Hua Jiang$^{5,6,7}$}
%\email{jhjiang3@ustc.edu.cn}
%\author{Lu-Jian Chen$^{4}$}
%\email{lujianchen@xmu.edu.cn}
%\author{Yan-qing Lu$^{8}$}
%\email{yqlu@nju.edu.cn}
%\author{Jin-hui Chen$^{1,2}$}
%\email{jimchen@xmu.edu.cn} 

\affiliation{$^{1}$Institute of Electromagnetics and Acoustics, Key Laboratory of Electromagnetic Wave Science and Detection Technology, Xiamen University, Xiamen 361005, China\\
$^2$Shenzhen Research Institute of Xiamen University, Shenzhen 518000, China\\
$^3$MinJiang Collaborative Center for Theoretical Physics, College of Physics and Electronic Information Engineering, Minjiang University, Fuzhou 350108, China\\
$^4$State Key Laboratory of Bioinspired Interfacial Materials Science, Suzhou Institute for Advanced Research, University of Science and Technology of China, Suzhou 215123, China\\
$^5$School of Physical Sciences, University of Science and Technology of China, Hefei 230026, China\\
$^6$College of Engineering and Applied Sciences, Nanjing University, Nanjing, 210023, China\\
$^7$Department of Electronic Engineering, Xiamen University, Xiamen 361005, China}

\date{\today}% It is always \today, today,

\begin{abstract}
%Branched flow—the formation of intense, filamentary channels of waves—emerges when light propagates through weakly disordered media. This striking phenomenon arises from fluctuating caustics due to random scattering, yet its inherent randomness has hindered controlled study and practical application. Achieving programmable control over optical branched flow remains an outstanding challenge.
\noindent We demonstrate deterministic control of branched flow of light using anisotropic nematic liquid crystals. By sculpting the director field via photoalignment, we create spatially programmable optical potentials that govern light scattering and propagation. This platform enables configurable, anisotropic branched flow of light and reveals a universal scaling law for its characteristic features, directly connecting disordered photonics with mesoscopic wave transport. Under extreme anisotropy, we observe a pronounced directional channeling effect, driven by anomalous symmetry-breaking velocity diffusion, which concentrates light propagation along preferential directions while suppressing transverse spreading. These findings establish a tunable material platform for harnessing branched flow of light, opening pathways toward on-chip photonic circuits that exploit disorder-guided transport, scattering-resilient endoscopic imaging, and adaptive optical interfaces in complex media.

\end{abstract}

\maketitle

%\tableofcontents

%\textbf{Introduction}

%\section*{Introduction}
The propagation of light through disordered media spans a rich spectrum of physical regimes, from ballistic transport in homogeneous media to fully diffusive spreading under strong multiple scattering. Between these extremes lies a fascinating intermediate regime where coherent interference in weakly correlated disordered potentials gives rise to a remarkable phenomenon: the branched flow of light~\cite{patsyk2020observation,patsyk2022incoherent}. Here, light spontaneously organizes into intense, filamentary channels that persistently branch during propagation, forming patterns reminiscent of river networks or lightning strikes. This behavior emerges universally across wave systems—from electrons and microwaves to water waves and light—when the correlation length of the disorder exceeds the wavelength~\cite{metzger2010universal,rotter2017light,garnier2025stochastic}. First discovered in two-dimensional electron gases~\cite{topinka2001coherent,jura2007unexpected,liu2013stability,maryenko2012branching}, branched flow has since been observed in optical systems such as thin soap films~\cite{patsyk2020observation} and tailored photonic platforms~\cite{chang2024electrical,liu2025}. However, experimental studies have remained largely observational, with limited ability to control or program the branching process. Crucially, key regimes such as anisotropic branched flow have remained virtually unexplored, while the lack of a programmable material platform has hindered the transition from fundamental observation to functional applications.

Theoretical insights suggest that anisotropy in the disordered potential offers a powerful tool to tailor branched flows~\cite{degueldre2017channeling,qincrossover}. For instance, anisotropic landscapes such as seabed topography can channel tsunami waves into extreme, rogue-wave-like amplitudes~\cite{degueldre2016random,gagnon2006multifractal}. Understanding and harnessing such anisotropic branching is thus essential for predicting and controlling extreme wave events in fields ranging from nonlinear optics to hydrodynamics and acoustics~\cite{dudley2014instabilities,berry2005tsunami,jiang2023branching}. Moreover, the ability to program branched flow of light—to dictate where, when, and how branches form—would open transformative opportunities in photonic technology. A reconfigurable platform capable of on-demand optical channeling could enable reprogrammable photonic circuits that guide light along dynamically tunable pathways, robust endoscopic imaging through scattering biological tissues, and adaptive optical interfaces that control light–matter interaction in complex media. Realizing such applications demands a material that combines precise spatial control over the disorder potential with notable stability---a combination that has remained elusive.

In this work, we introduce a programmable material platform that achieves deterministic control over branched light flow by leveraging the tunable anisotropy and microscale patterning capability of nematic liquid crystals (NLCs)~\cite{xu2024modulating,ma2022self,zhao2025space,li2018electrically,wang2024moire,wang2024vectorial}. Using high-resolution photoalignment, we engineer spatially correlated disorder with continuous and adjustable anisotropy, creating optical potentials that are both smooth and fully reconfigurable. This approach transforms branched flow from a passively observed phenomenon into an actively controllable wave-transport effect. With this platform, we experimentally demonstrate on-demand customization of branching morphology, validate a universal scaling law for anisotropic branched flow, and uncover a previously unreported regime of anomalous velocity diffusion under extreme anisotropy. The latter leads to a pronounced directional channeling effect, effectively guiding light along designer pathways without the need for conventional waveguides. Our work not only provides a fundamental breakthrough in the control of wave propagation in disordered media, but also establishes a versatile platform for applications in on-chip photonics, biomedical imaging, and adaptive optical systems.

%Our work represents a fundamental breakthrough in wave control within disordered media, bridging the gap between ballistic and diffusive transport. It establishes a versatile experimental framework for studying wave branching, scaling, and localization in tailored potentials. Moreover, it directly addresses practical needs in on-chip photonics---where reconfigurable optical routing is sought after---and biomedical optics---where controlled light delivery through scattering tissue remains a grand challenge. By providing a pathway to programmable branched flow, this platform opens new frontiers in disordered photonics, from reconfigurable integrated optical devices and enhanced deep-tissue imaging to novel schemes for optical manipulation and sensing in complex environments.

\begin{figure}[tb]%
\centering
\includegraphics[width=0.48\textwidth]{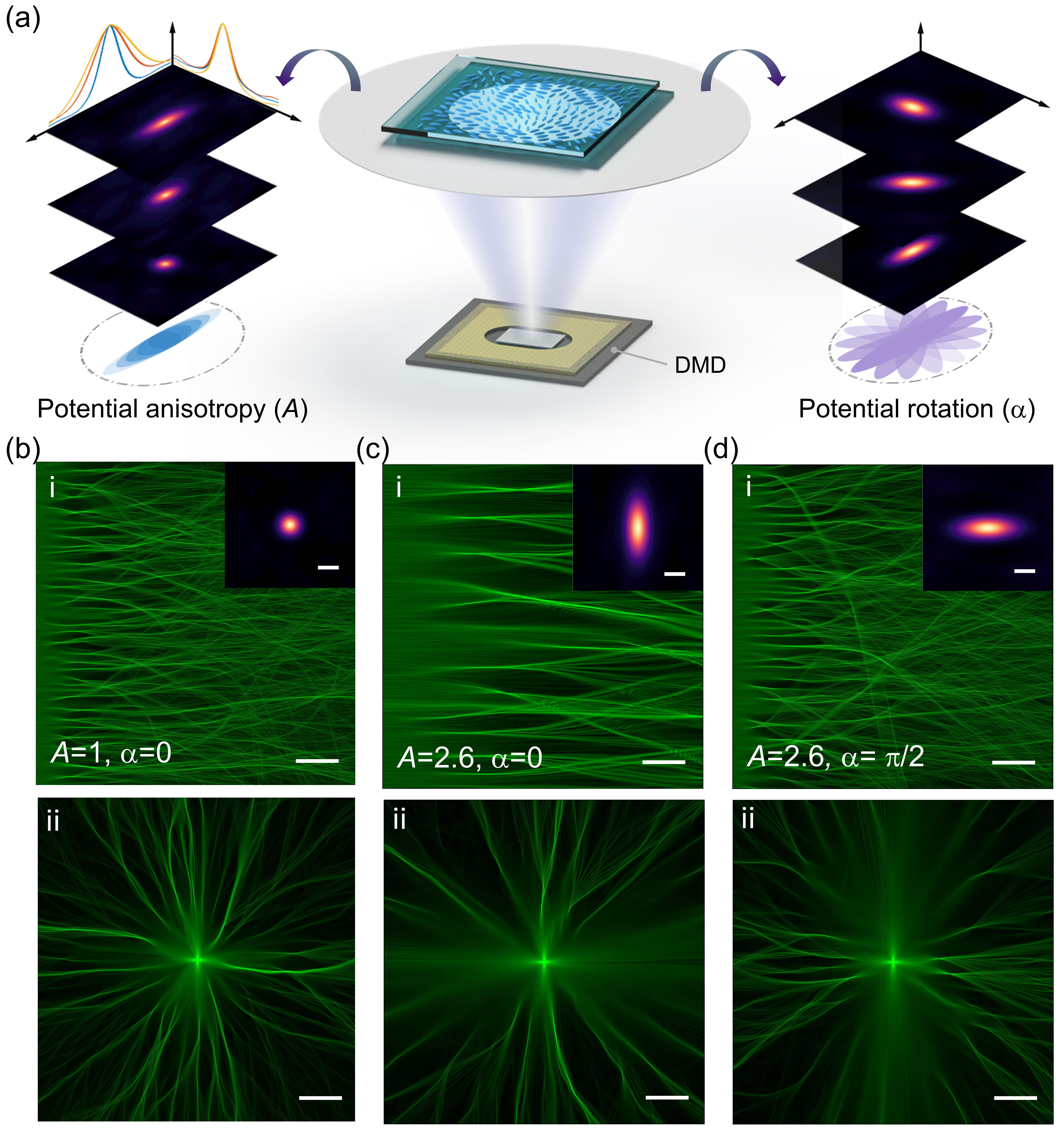}
\caption{Programmable optical branched flow in nematic liquid crystals (NLC). (a) Tailoring disordered potential of photo-alignment NLC film via potential anisotropy and rotation for controlling anisotropic branched flow. DMD: digital micro-mirror device. %The potential anisotropy ($A$) is defined as $A=l_y/l_x$, where $l_x$ and $l_y$ are correlation length of optical potential in the two orthogonal directions. The potential rotation angle ($\alpha$) is obtained by coordinate transformation of disordered potential.
%The effective optical potential of NLC film is determined by their director distributions, which is fabricated based on the digital micro-mirror device (DMD) microlithography system. The pixelated grayscale patterns are exposed to the liquid crystal film induced by the surface photoalignment layer. 
%The insets: (i) designed two-dimensional (2D) topographic maps of NLC director distributions with varying potential anisotropy; (ii) 2D topographic maps of NLC director distributions with different potential rotation. DMD: digital micro-mirror device. %The scale bars are 200 $\upmu$m.
(b-d) The simulated optical branched flow fields in the anisotropic disordered NLC film with different potential anisotropy ($A$) and rotation ($\alpha$). %The scale bars are 200 $\upmu$m.
The panels (i) are the simulated branched flow fields with a plane-wave excitation, where the insets show the correlation function of the corresponding disordered potential. The panels (ii) are the simulated branched flow fields via a point-source excitation. %The light channeling effect is generally along the long-axis direction of disordered anisotropic potential.
Here the correlation length $l_x$ is set constant as 15 $\upmu$m. %the pixel size of photopatterned NLC is xx $\upmu$m according to the experimental conditions.
All the scale bars are 200 $\upmu$m.}
\label{fig1}
\end{figure}

%\noindent{\textbf{Results}}
%\section*{Results}

%\noindent{\textbf{Theoretical design of anisotropic disordered potential in an NLC film}}
In our platform, the NLC film serves as a slab waveguide that effectively supports guided light propagation and enables us to study branched flow of light. The propagation of the guided extraordinary light wave ($e$-wave) is governed by the spatial distribution of the director field $\hat{\textbf{\textit{n}}}(x,y)$ of the liquid crystal molecules, which determines the effective refractive index distribution $n_{\rm{neff}}(x,y)$ felt by the $e$-wave~\cite{khoo1993optics}. The thickness of the implemented NLC film (20~$\upmu$m) is much larger than the wavelength of the laser light (532~nm) used in our experiments, and thus the waveguide-induced effective index change is neglected~\cite{chang2024electrical}. The equation governing the propagation of a monochromatic light in the NLC film is approximately 2D Helmholtz equation, which is written in a form resembling the time-independent Schr\"{o}dinger equation,
\begin{equation}
    -\nabla^2\psi+V(x,y)\psi=k_0^2\bar{n}^2\psi,
\end{equation}
where $\psi$ is the light wave amplitude and $V(x,y)$ is the effective optical potential given by (see Supplementary Material for details)\cite{patsyk2022incoherent,chang2024electrical,choate2015optimization,supplemental}:
\begin{equation}
V(x,y)=k_0^2(\bar{n}^2-n_{\rm{eff}}(x,y)^2),
\end{equation}
where $k_0$ is the wavevector of the laser light in vacuum, $\bar{n}=\sqrt{\left\langle{n_\mathrm{eff}^2(x,y)}\right\rangle}$ is the (root-mean-square) averaged refractive index in the observation area of the NLC film. The average optical potential is $\left\langle{V(x,y)}\right\rangle\equiv 0$, indicating purely the effect of the spatial fluctuation of potentials. Physically, $V(x,y)$ gives a smooth, spatially varying disordered potential that scatters the light and results in the branched flows. The spatial correlation of the optical potential $c(x',y')=\left\langle{V(x',y')V(0)}\right\rangle$ is a quantity that governs the key properties of the branched flow of light~\cite{patsyk2020observation,kaplan2002statistics,metzger2010universal}. For instance, in an isotropic medium, the correlation function of the disordered optical potential always falls into the form, $c(x',y')=\epsilon^2f((x'^2+y'^2)/l_c^2)$, where $f(\cdot)$ is a dimensionless form function. The first-branch distance $d_f$\textemdash the characteristic length scale of branched flows\textemdash is defined as the statistically averaged distance from the plane-wave source to the first caustic focusing positions. $d_f$ is determined by the correlation length $l_c$ and relative strength $\epsilon$ of the disordered potential.%~\cite{degueldre2017channeling}. 

\begin{figure}[t]%
\centering
\includegraphics[width=0.48\textwidth]{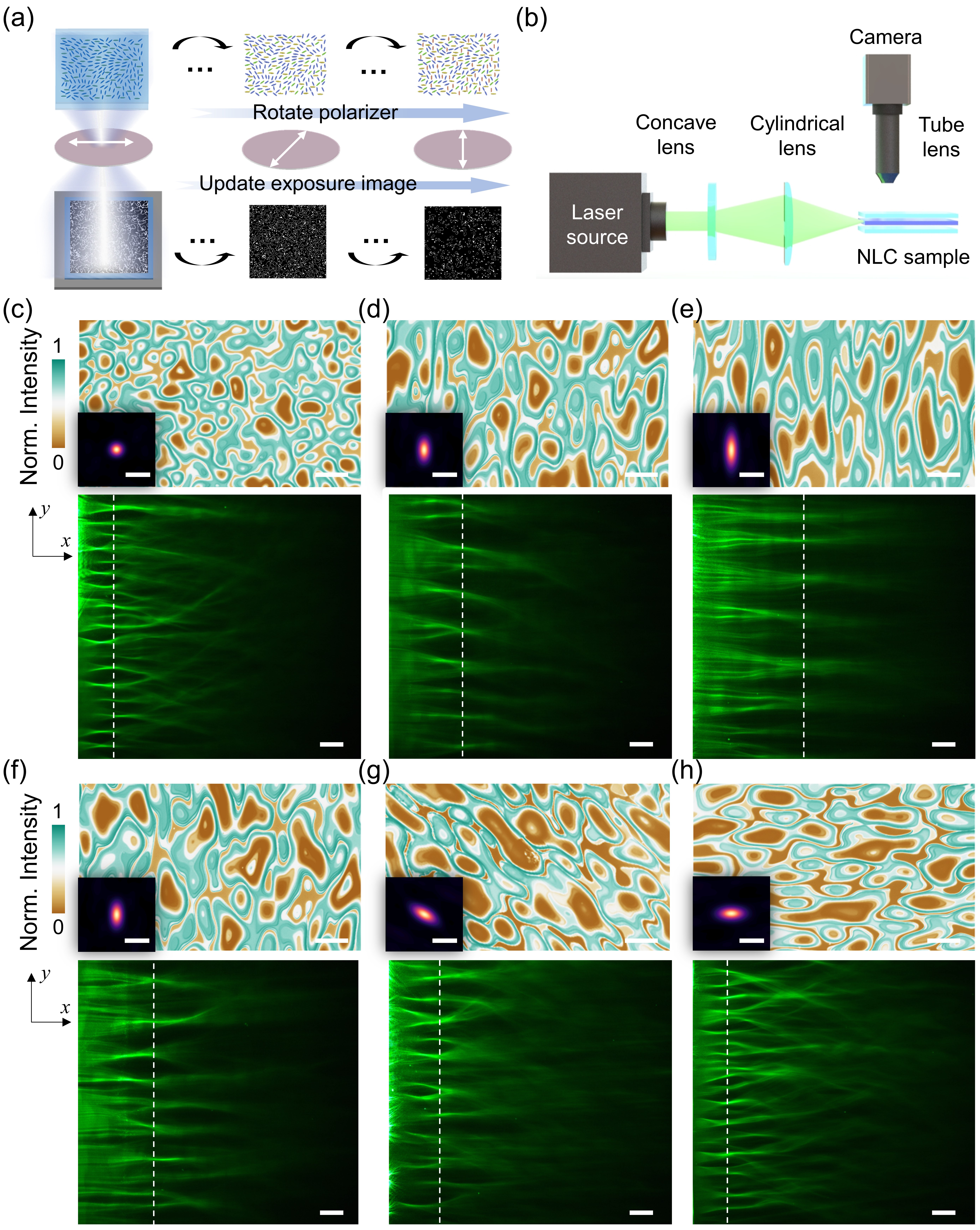}
\caption{%Scaling law between the length of the first branch and the spatial anisotropy. 
Experimental observation of anisotropic branched flow with controllable first-branch distance. 
(a) Schematic diagram of photo-patterning NLC films with designed disordered director distributions. %The NLC director distributions are obtained by sequentially updating of exposure image and the rotation angle of a linear polarizer. %, both of which are synchronously controlled by a computer.
%(b) The quantified disordered parameters ($\epsilon, A, \alpha$) of correlated potential across the fabricated NLC samples. 
(b) Experimental setup for excitation and characterization of branched flow in an NLC sample. (c-h) Cross-polarized microscope images (upper panels) of fabricated NLC films with varying potential anisotropy $A$ and potential rotation angle $\alpha$, and their correspondingly measured profiles of the propagating field (lower panels). 
In (c-e) the disordered potentials are of the same $\alpha=0$, and different $A$-values: (c) $A=1.0$, (d) $A=1.8$. (e) $A=2.6$. In (f-h) the disordered potentials are of the same $A=1.8$, and different $\alpha-$values: (f) $\alpha=0$, (g) $\alpha=\pi/4$, (h) $\alpha=\pi/2$. %The quasi-plane-wave incident light is propagating along the $x$-direction with a linear polarization along the $y$-direction.
The insets in the upper panels of (c-h) are the autocorrelation function of the disordered optical potential. The white dashed line in lower panels of (c-h) indicates the measured first-branch distance.  All the scale bar is 100 $\upmu$m. } % 确认下白色虚线计算方法--实验上计算闪烁指数（约十个位置的平均）峰值的位置
\label{fig2}
\end{figure}

Theoretically, the position of the first caustic, or the first-branch distance, obeys a generalized relation in anisotropic disordered potentials (see Supplementary Material for details)~\cite{degueldre2017channeling,supplemental}:
\begin{equation}
d_f \propto \epsilon^{-2/3} \left(l_y^4/{l_x} \right)^{1/3}=\epsilon^{-2/3}A^{4/3}l_x,
\end{equation}
where $l_x$ and $l_y$ are the correlation length along the longitudinal and transverse directions, respectively, 
$A=l_y/l_x$ is the potential anisotropy. In a more general anisotropic disordered medium, where the principal axis is tilted by an angle $\alpha$ relative to the $x$-direction, the angular dependence of the first-branch distance can be obtained:
\begin{equation}
d_f \propto {\epsilon}^{-2/3}A^{4/3}l_x(\rm{cos}^2\alpha+\rm{sin}^2\alpha/\textit{A}^2)^{5/6}
\end{equation}
In the experiment, by tailoring director distribution of NLC films to construct anisotropic disordered potentials using the photoalignment technique (Fig.~\ref{fig1}a)~\cite{wei2020liquid,wei2016liquid,wang2024vectorial,meng2022topological}, the branched flow of light can be programmed and controlled on demand.

%We thus can design the disordered potential by tuning the potential anisotropy $A$ and the rotation angle $\alpha$, as shown in Fig.~\ref{fig1}a, which can efficiently control the branched flows.

The disordered optical potential is generated by convolving a 2D random matrix with Gaussian white noise and the target anisotropic correlation function (see Supplementary Material for details~\cite{supplemental}). Since branched flow arises from random small-angle deflection events induced by a weakly correlated potential, the input light source is required to be weakly localized in phase space, such as a plane-wave source~\cite{heller2021branched}. The simulated branched light fields under plane-wave illumination (Figs.~\ref{fig1}b-d, panels i) align well with the theoretical scaling law for the first-branch distance $d_f$. For example, $d_f$ for $(A=1,\alpha=0)$ is smaller than for $(A=2.6,\alpha=0)$, while $d_f$ for $(A=2.6,\alpha=0)$ exceeds that for $(A=2.6,\alpha=\pi/2)$. We also simulate light propagation with a point-source excitation (panels ii of Figs.~\ref{fig1}b-d), which reveals a strong polar angle dependence of the wave caustics distribution in anisotropic potentials. Furthermore, a self-channeling effect emerges along a direction where the correlation of the disorder potential is strong, consistent with the previous theoretical predictions~\cite{degueldre2017channeling}. These properties are examined in detail in the following sections.

\begin{figure}[tb]%
\centering
\includegraphics[width=0.42\textwidth]{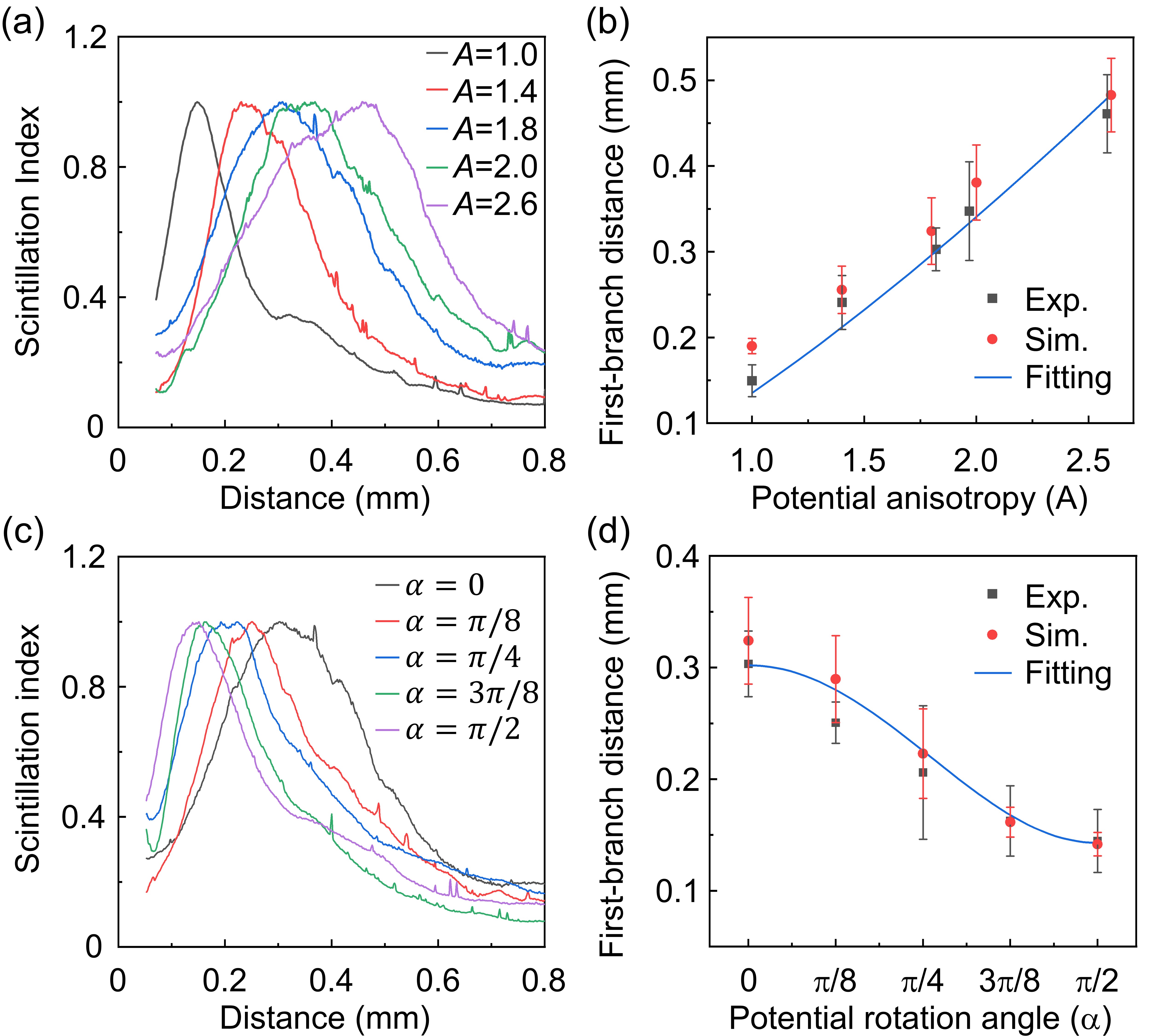}
\caption{Quantitative measurement of the first-branch distance on potential anisotropy and potential rotation. (a) The shift of scintillation index curves on varying potential anisotropy. (b) Experimental and simulated results of first-branch distance depending on potential anisotropy. %The fitting curve is ${1.115\epsilon}^{-2/3}A^{4/3}l_x$, where $\epsilon=0.048$, $l_x=16~\upmu$m. %The error bar is 98$\%$ peak bandwidth of the normalized scintillation index curve.
(c) The shift of scintillation index curves on varying potential rotation angle. The potential anisotropy is set as 1.8. (d) Experimental and simulated results of first-branch distance depending on potential rotation angle. %The fitting curve is ${1.2\epsilon}^{-2/3}A^{4/3}l_x(\rm{cos}^2\alpha+\rm{sin}^2\alpha$/$A^2$). 
The error bars in (b,d) is bandwidth of 98$\%$ peak value of the normalized scintillation index curve.}
\label{fig3}
\end{figure} 
%\noindent{\textbf{Experimental study of anisotropic branched flow of light}}

To realize an effective optical potential for anisotropic branched flow, we first fabricate inhomogeneously aligned NLC films. The director distribution is digitized into 18 gray levels, corresponding to NLC director angles ranging from 0 to $17\pi/18$ radians in steps of $\pi$/18 (see Supplementary Material for details~\cite{supplemental}). This fine discretizations of NLC directors suppress the externally angular dispersion effect of $e$-wave in the NLCs (see Supplementary Material for details~\cite{supplemental}). The designed NLC films are fabricated using the photoalignment technique, where the NLC director profile is created through a multi-step partial ultraviolet exposure combined with synchronous rotation of the linear polarizer as illustrated in Fig.~\ref{fig2}a.
Experimentally, we primarily vary the anisotropy $A$ and the rotation angle $\alpha$ of the disordered potential. The resulting parameter sets $(\epsilon, A, \alpha)$ for the fabricated films are summarized (see Supplementary Material for details~\cite{supplemental}). 
%in Fig.~\ref{fig2}b. 
The optical potential strength $\epsilon$ is approximately 0.05, as determined from the birefringence of the E7 mixture ($\Delta n=0.22$ in the visible range)~\cite{chang2024electrical,wei2020liquid}. To characterize the emergence of the first-branch distance, we couple a quasi-plane-wave light into the NLC film and directly observe the in-plane propagating light field using a home-made optical microscope (Fig.~\ref{fig2}b).
This far-field measurement scheme is enabled by the intrinsic weak light scattering out of the NLC film plane~\cite{khoo1993optics,bender2022coherent}.

The emergent disordered textures are verified with cross-polarized optical microscopy (upper panels of Figs.~2c–h). Furthermore, spatial variations in correlation length and potential strength are characterized in different sections within a single NLC film, revealing good self-similarity for the observation of branched flow of light (see Supplementary Material for details~\cite{supplemental}). From the lower panels of Figs.~\ref{fig2}(c-e), the incident light evolves from an initially uniform intensity distribution to pronounced caustics at a characteristic position $d_f$ (approximately the first-branch distance~\cite{patsyk2020observation,patsyk2022incoherent}). The caustic distance increases monotonously with the anisotropy of the disordered potential due to the increase in the correlation length along the $y$-direction. In contrast, at a fixed anisotropy ($A=1.8$), the caustic distance decreases with increasing rotation angle in the range of $0\sim \pi/2$ (Figs.~\ref{fig2}f-h). This pronounced dependence on propagation angle originates from the variations in the effective transverse correlation length.

Figure \ref{fig3}a presents the measured scintillation index curves for five groups of samples with various potential anisotropy. Each curve represents an ensemble average over approximately ten independent realizations, in which the NLC films are designed and fabricated to have nearly identical statistical parameters ($\epsilon, A, \alpha$). This is enabled by the precisely engineered disordered potential via high-accuracy photoalignment. %给出一个参数下，制备的多个样品表征图，体现不同样品统计参数一致性%,
The high programmability of NLC platform is essential for achieving meaningful ensemble averaging. The scintillation index is defined as the normalized variance of the lateral light intensity: $S(x)=\left\langle{I^2(x)}\right\rangle/\left\langle{I(x)}\right\rangle^2-1$, where averaging is performed over the transverse $y$-direction and across different realizations~\cite{barkhofen2013experimental,chang2024electrical,patsyk2020observation}. Physically, $S(x)$ reaches its maximum when the transverse intensity fluctuations are strongest, marking the onset of the branching. The experimentally measured first-branch distance $d_f$ shows quantitative agreement with both the analytical scaling law from equation (3) and numerical simulations (see Supplementary Material for details~\cite{supplemental}). Furthermore, the measured $d_f$ strongly depends on the potential rotation angle, closely following the theoretical predictions (Figs.~\ref{fig3}c-d). Counterintuitively, in the polar representation $(d_f,\alpha)$, the angular dependence does not exhibit a simple ellipse but instead shows a distinct ``peanut" shape (see Supplementary Material for details~\cite{supplemental}). These results provide a direct experimental discovery of the universal scaling law for anisotropic branched flow.  %$\epsilon^{-2/3}\left(l_y^4/l_x\right)^{1/3}$, reaches its maximum value, and then slowly decreases with a long tail.
%The enhanced spatial anisotropy leads to a trend of overall backward shift in the peak position of scintillation index, thus experimentally revealing the inherent scaling laws of various spatially anisotropic disordered potential landscapes. The peak positions of scintillation indices the white dashed lines in Fig. 2b. 
%\noindent{\textbf{Observation of light channeling effect in extremely anisotropic potential}
\begin{figure}[tb]
\centering
\includegraphics[width=0.48\textwidth]{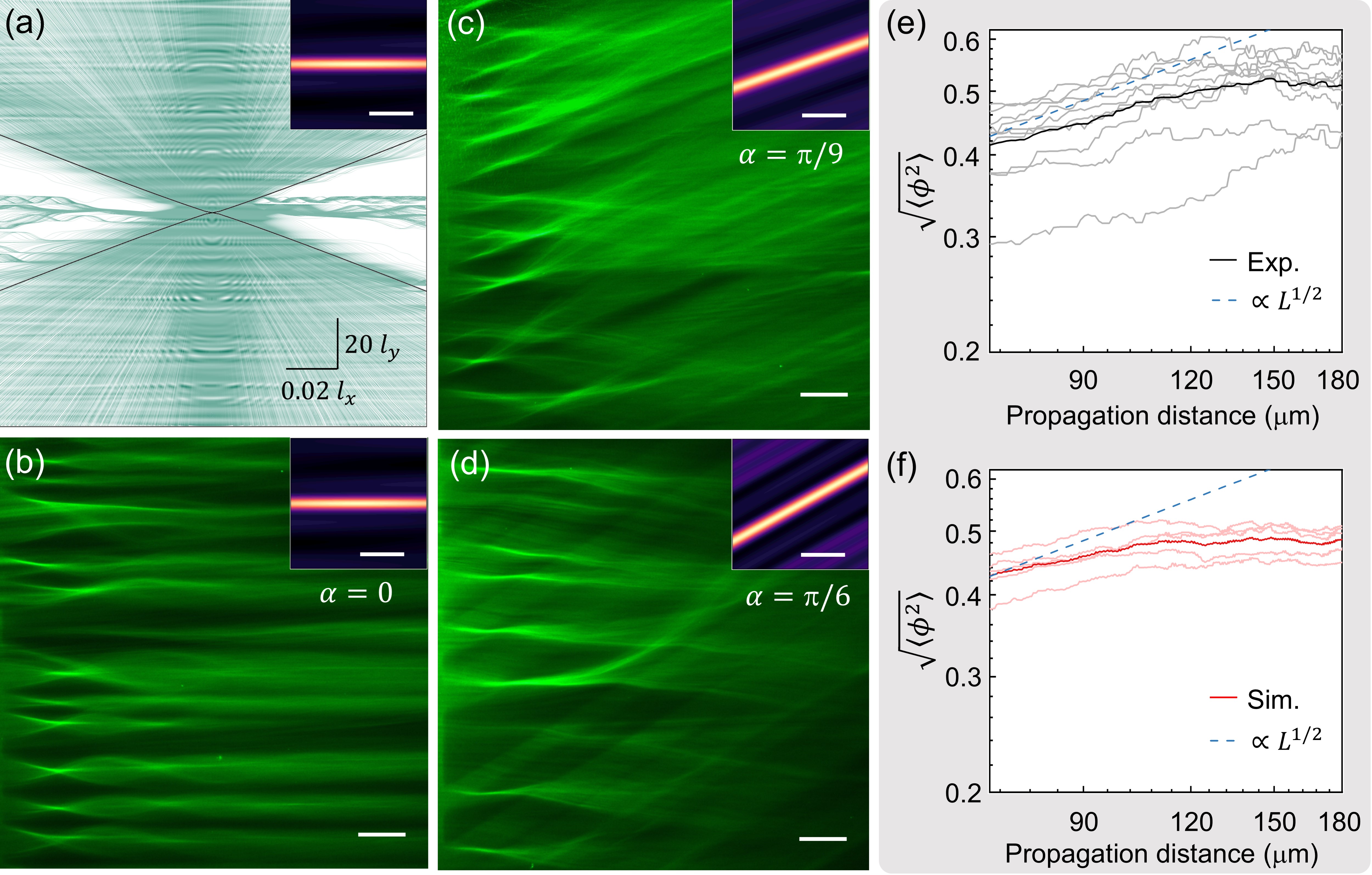}
\caption{Observation of light channeling effect in extreme anisotropic disordered potential. (a) Simulated branched flow field in spatially anisotropic disordered NLC films, with anisotropy of $A=1/1000$. Here $l_y$ is set as 16 $\upmu$m. The scale bar is represented by the correlation lengths. %The color map visualizes the spatial distribution of trajectory density from a central point source.
(b-d) The microscope images show the branching of a wide elliptical laser beam propagating through extreme anisotropic potentials ($A=1/1000$) at spatial orientation angles of (b) 0, (c) $\pi/9$, and (d) $\pi/6$. %highlighting the light localization effect in anisotropic potential. 
The insets in (a-d) show the autocorrelation function of the disordered potential. The scale bar is 100 $\upmu$m. (e-f) The standard deviation of the velocity angle for propagating light in an extremely anisotropic potential ($A=1/1000$): (e) experimental results, (f) simulated results. %Ensemble averaging over ten measurements (five simulations) yields the black (red) curve.
The undertint curves are examples of realizations as an illustration for the typical variation in individual realizations of the disordered NLCs. The dashed-blue lines in (e-f) are proportional to $L^{1/2}$ for normal diffusion process, where $L$ is the propagation distance.}
\label{fig4}
\end{figure}

In an isotropic disordered potential, the light field landscape excited by a point light source exhibits a branched caustic pattern that is essentially independent of the propagation angle (see Supplementary Material for details~\cite{supplemental}).
%(Fig.~\ref{fig4}a).
In contrast, anisotropic disordered potentials tend to concentrate optical caustics along the longer correlation length (panels ii of Figs.~\ref{fig1}c-d)~\cite{degueldre2017channeling}. In experiment, when excited with a narrow Gaussian beam, the self-channeling effect is enhanced with increasing potential anisotropy (see Supplementary Material for details~\cite{supplemental}). This effect is particularly pronounced in extremely spatially anisotropic media. For example, at $l_x:l_y=1000$ or $A$=1/1000, (Fig.~\ref{fig4}a), the light field can be strongly localized around the $x$-axis, corresponding to the direction of the longer correlation length. Therefore, this light channeling regime can be shifted by simply rotating the potential (see Supplementary Material for details~\cite{supplemental}). 

For such an extremely anisotropic case ($A\ll1$ or $l_y\ll{l_x}$), the disordered potential distribution is only weakly dependent on the $x$-coordinate. Therefore, the energy exchange can be approximated as occurring solely in the the $y$-direction, i.e., $E_{total}=E_y+U(y)$. When all kinetic energy is converted into potential energy, i.e., $U=v_0^2\rm{sin}^2\beta/2$, the transverse component of the light's trajectory reverses direction, while the longitudinal ($x$) propagation remains monotonic. According to the Piterbarg's theorem~\cite{krogstad2004spatial,degueldre2017channeling}, the maximum lateral excursion $y_m$ is related to the peak potential via $U_{max}=\epsilon\sqrt{2{\rm{ln}}(y_m/l_c)+2}$. Along with $U_{max}=v_0^2\rm{sin}^2\beta/2$, %${\rm{tan}}\beta=\frac{y_m}{x_m}$, 
the asymptotic curve for optical localization is obtained, shown as the black line in Fig.~\ref{fig4}a. Most of the light caustics are confined within the cone-like region, producing a nontrivial optical channeling effect.%没有具体的临界角，而是根据公式可以画出（xm，ym）这条线

To probe light localization, we imprint the designed, extremely anisotropic disordered potential onto the NLC director field via photo-alignment and launch a quasi-plane wave into samples with varying potential rotation angles $\alpha$. For small rotation angles $\alpha\leq \pi/9$, the resulting caustics remain tightly confined along the principal axis of the potential (Figs.~\ref{fig4}b–c), exhibiting a pronounced waveguiding-like behavior (see Supplementary Material~\cite{supplemental} for additional data). In contrast, when $\alpha$ reaches $\pi/6$---slightly exceeding the theoretical asymptotic angle---the emergent caustics no longer align with the potential’s principal axis. Instead, they persist along the initial propagation direction (Fig.~\ref{fig4}d). These experimental observations are in excellent agreement with the theoretical profile shown in Fig.~\ref{fig4}a.

To elucidate how spatial anisotropy governs light localization, we analyze the diffusion of velocity angles during the light propagation in an extremely spatially anisotropic potential ($A=1/1000$, $\alpha=0$). % In a pure diffusive process, the probability density of velocity angle follows a Gaussian distribution $p(\phi,x)=\frac{1}{\sqrt{4\pi\sigma^2x}}\rm{exp}(-\frac{\phi^2}{4\sigma^2\textit{x}})$. %where $\sigma^2(\alpha)$ is the diffusion constant, reflecting the diffusion rate at each angle. sigma之前定义过 The standard deviation of velocity angle $\sqrt{\left\langle{\phi^2}\right\rangle(x)}=\sqrt{2\sigma^2x}$ scales as $x^{1/2}$.  %这部分放到SI
We experimentally and numerically evaluate the dependence of the velocity-angle standard deviation, $\sqrt{\left\langle{\phi^2}\right\rangle(x)}$, on the propagation distance $L$, as shown in Fig. \ref{fig4}e-f (see Supplementary Material for details~\cite{supplemental}). The extracted growth exponent $s$ in $\sqrt{\left\langle{\phi^2}\right\rangle(x)}\propto L^s$ is smaller than 1/2, indicating a high probability of light remaining localized along the direction of longer correlation length, exhibiting a sub-diffusive transport regime (see Supplementary Material for details~\cite{supplemental}).
%To elucidate how spatial anisotropy governs light localization, we analyze the diffusion of velocity angles during the light propagation. The probability density of velocity angle in pure diffusion process satisfies Gaussian distribution $p(\phi,x)=\frac{1}{\sqrt{4\pi\sigma^2x}}\rm{exp}(-\frac{\phi^2}{4\sigma^2x})$, where $\sigma^2(\alpha)$ is the diffusion constant, reflecting the diffusion rate at each angle. The standard deviation of velocity angle $\sqrt{\left\langle{\phi^2}\right\rangle(x)}=\sqrt{2\sigma^2x}$ increases exponentially with propagation distance by 1/2. We statistically analyzed the variation of velocity angle standard deviation $\sqrt{\left\langle{\phi^2}\right\rangle(x)}$ with propagation distance in extreme spatial anisotropic potentials under the initial condition of plane waves with a correlation length direction of $\alpha=0$. Its growth index is lower than 1/2, indicating a high probability that light is localized in the direction of the longer axis of the correlation function and exhibits a sub-diffusion state (see Supplementary Note S7 and Fig. S15).
~~
%\noindent{\textbf{Conclusion and outlook}}

In summary, we have demonstrated deterministic control over anisotropic branched flow of light by engineering programmable, spatially correlated disorder in nematic liquid crystals. This represents a fundamental advance: transforming branched flow from a passively observed wave phenomenon into an actively sculpted and reconfigurable transport regime. Our platform—enabled by photo-alignment techniques that offer micron-scale precision over the director field—allows the optical potential landscape to be tailored in both strength and anisotropy, providing unprecedented command over caustic formation and branching morphology. The experimental validation of a universal scaling law for the first-branch distance in anisotropic media underscores the role of engineered disorder as a governing factor in wave transport, bridging ballistic and diffusive propagation. Beyond scaling, we uncover that extreme spatial anisotropy triggers anomalous velocity diffusion, a symmetry-breaking process that dynamically confines light along preferential directions and leads to pronounced, tunable optical channeling. These findings reshape the understanding of wave scattering in correlated disorder, establishing anisotropy not merely as a parameter but as a design degree of freedom for controlling light in complex media.

This work is supported by National Key Research and Development Program of China (2023YFA1407104 and 2022YFA1405000), National Natural Science Foundation of China (12125504, 12274357, and 62475223), Shenzhen Science and Technology Program (JCYJ20240813145614019), Fujian Provincial Natural Science Foundation of China (2023J06011), and the ``Hundred Talents Program'' of the Chinese Academy of Sciences.

%\noindent{\textbf{Author contributions}}

%S.-s.C. fabricated the samples with the assistance of L.-J.C.; S.-s.C. performed the experiments; S.-s.C. and J.-h.C. built the theoretical model with the assistance of D.X.; Z.-h.Z, L.-W.W., J.-H.J., and Y.-q.L; S.-s.C. and J.-h.C. analyzed the results and drew the figures. J.-h.C. conceived the idea and co-supervised the project with J.-H.J., L.-J.C and Y.-q.L. All authors contributed to the discussions of the results and manuscript preparation. S.-s.C., J.-H.J. and J.-h.C. wrote the manuscript and the supplementary information.
\bibliography{Ref}

@PREAMBLE{
 "\providecommand{\noopsort}[1]{}" 
 # "\providecommand{\singleletter}[1]{#1}%" 
}

@article{heller2021branched,
  title={Branched flow},
  author={Heller, Eric J and Fleischmann, Ragnar and Kramer, Tobias},
  journal={Phys. Today},
  volume={74},
  number={12},
  pages={44--51},
  year={2021},
  publisher={AIP Publishing}
}

@article{topinka2001coherent,
  title={Coherent branched flow in a two-dimensional electron gas},
  author={Topinka, MA and LeRoy, BJ and Westervelt, RM and Shaw, SEJ and Fleischmann, R and Heller, EJ and Maranowski, KD and Gossard, AC},
  journal={Nature},
  volume={410},
  number={6825},
  pages={183--186},
  year={2001},
  publisher={Nature Publishing Group}
}

@article{liu2013stability,
  title={Stability of branched flow from a quantum point contact},
  author={Liu, Bo and Heller, Eric J},
  journal={Phys. Rev. Lett.},
  volume={111},
  number={23},
  pages={236804},
  year={2013},
  publisher={APS}
}

@article{jura2007unexpected,
  title={Unexpected features of branched flow through high-mobility two-dimensional electron gases},
  author={Jura, MP and Topinka, MA and Urban, L and Yazdani, A and Shtrikman, Hadas and Pfeiffer, LN and West, KW and Goldhaber-Gordon, D},
  journal={Nat. Phys.},
  volume={3},
  number={12},
  pages={841--845},
  year={2007},
  publisher={Nature Publishing Group}
}

@article{maryenko2012branching,
  title={How branching can change the conductance of ballistic semiconductor devices},
  author={Maryenko, Denis and Ospald, F and Klitzing, K v and Smet, JH and Metzger, JJ and Fleischmann, R and Geisel, T and Umansky, V},
  journal={Phys. Rev. B},
  volume={85},
  number={19},
  pages={195329},
  year={2012},
  publisher={APS}
}

@article{degueldre2016random,
  title={Random focusing of tsunami waves},
  author={Degueldre, Henri and Metzger, Jakob J and Geisel, Theo and Fleischmann, Ragnar},
  journal={Nat. Phys.},
  volume={12},
  number={3},
  pages={259--262},
  year={2016},
  publisher={Nature Publishing Group}
}

@article{barkhofen2013experimental,
  title={Experimental observation of a fundamental length scale of waves in random media},
  author={Barkhofen, Sonja and Metzger, Jakob J and Fleischmann, Ragnar and Kuhl, Ulrich and St{\"o}ckmann, H-J},
  journal={Phys. Rev. Lett.},
  volume={111},
  number={18},
  pages={183902},
  year={2013},
  publisher={APS}
}

@article{patsyk2020observation,
  title={Observation of branched flow of light},
  author={Patsyk, Anatoly and Sivan, Uri and Segev, Mordechai and Bandres, Miguel A},
  journal={Nature},
  volume={583},
  number={7814},
  pages={60--65},
  year={2020},
  publisher={Nature Publishing Group}
}

@article{patsyk2022incoherent,
  title={Incoherent branched flow of light},
  author={Patsyk, Anatoly and Sharabi, Yonatan and Sivan, Uri and Segev, Mordechai},
  journal={Phys. Rev. X},
  volume={12},
  number={2},
  pages={021007},
  year={2022},
  publisher={APS}
}

@article{gagnon2006multifractal,
  title={Multifractal earth topography},
  author={Gagnon, J-S and Lovejoy, S and Schertzer, D},
  journal={Nonlin. Processes Geophys.},
  volume={13},
  number={5},
  pages={541--570},
  year={2006},
  publisher={Copernicus Publications G{\"o}ttingen, Germany}
}

@article{degueldre2017channeling,
  title={Channeling of branched flow in weakly scattering anisotropic media},
  author={Degueldre, Henri and Metzger, Jakob J and Schultheis, Erik and Fleischmann, Ragnar},
  journal={Phys. Rev. Lett.},
  volume={118},
  number={2},
  pages={024301},
  year={2017},
  publisher={APS}
}

@article{chang2024electrical,
  title={Electrical tuning of branched flow of light},
  author={Chang, Shan-shan and Wu, Ke-Hui and Liu, Si-jia and Lin, Zhi-Kang and Wu, Jin-bing and Ge, Shi-jun and Chen, Lu-Jian and Chen, Peng and Hu, Wei and Xu, Yadong and others},
  journal={Nat. Commun.},
  volume={15},
  number={1},
  pages={197},
  year={2024},
  publisher={Nature Publishing Group UK London}
}

@article{wei2020liquid,
  title={Liquid-Crystal-Mediated Active Waveguides toward Programmable Integrated Optics},
  author={Wei, Ting and Chen, Peng and Tang, Ming-Jie and Wu, Guang-Xing and Chen, Zhao-Xian and Shen, Zhi-Xiong and Ge, Shi-Jun and Xu, Fei and Hu, Wei and Lu, Yan-Qing},
  journal={Adv. Opt. Mater.},
  volume={8},
  number={10},
  pages={1902033},
  year={2020},
  publisher={Wiley Online Library}
}

@article{wei2016liquid,
  title={Liquid crystal depolarizer based on photoalignment technology},
  author={Wei, Bing-Yan and Chen, Peng and Ge, Shi-Jun and Zhang, Li-Chao and Hu, Wei and Lu, Yan-Qing},
  journal={Photon. Res.},
  volume={4},
  number={2},
  pages={70--73},
  year={2016},
  publisher={Chinese Laser Press and Optical Society of America}
}

@article{wang2024vectorial,
  title={Vectorial liquid-crystal holography},
  author={Wang, Ze-Yu and Zhou, Zhou and Zhang, Han and Wei, Yang and Yu, Hong-Guan and Hu, Wei and Chen, Wei and Dai, Hai-Tao and Ma, Ling-Ling and Qiu, Cheng-Wei and others},
  journal={eLight},
  volume={4},
  number={1},
  pages={5},
  year={2024},
  publisher={Springer}
}

@article{kaplan2002statistics,
  title={Statistics of branched flow in a weak correlated random potential},
  author={Kaplan, Lev},
  journal={Phys. Rev. Lett.},
  volume={89},
  number={18},
  pages={184103},
  year={2002},
  publisher={APS}
}

@article{metzger2010universal,
  title={Universal statistics of branched flows},
  author={Metzger, Jakob J and Fleischmann, Ragnar and Geisel, Theo},
  journal={Phys. Rev. Lett.},
  volume={105},
  number={2},
  pages={020601},
  year={2010},
  publisher={APS}
}

@article{choate2015optimization,
  title={Optimization of electromagnetic wave propagation through a liquid crystal layer},
  author={Choate, Eric P and Zhou, Hong},
  journal={Discrete Contin. Dyn. Syst.},
  volume={8},
  number={2},
  pages={303},
  year={2015},
  publisher={American Institute of Mathematical Sciences}
}

@inproceedings{krogstad2004spatial,
  title={Spatial extreme value analysis of nonlinear simulations of random surface waves},
  author={Krogstad, Harald E and Liu, Jingdong and Socquet-Juglard, Herve and Dysthe, Kristian B and Trulsen, Karsten},
  booktitle={International Conference on Offshore Mechanics and Arctic Engineering},
  volume={37440},
  pages={285--295},
  year={2004}
}

@book{khoo1993optics,
  title={Optics and nonlinear optics of liquid crystals},
  author={Khoo, Iam-Choon and Wu, Shin-Tson},
  volume={1},
  year={1993},
  address={Singapore},
  publisher={World Scientific}
}

@article{rotter2017light,
  title={Light fields in complex media: Mesoscopic scattering meets wave control},
  author={Rotter, Stefan and Gigan, Sylvain},
  journal={Rev. Mod. Phys.},
  volume={89},
  number={1},
  pages={015005},
  year={2017},
  publisher={APS}
}

@article{meng2022topological,
  title={Topological steering of light by nematic vortices and analogy to cosmic strings},
  author={Meng, Cuiling and Wu, Jin-Sheng and Smalyukh, Ivan I},
  journal={Nat. Mater.},
  volume={22},
  pages={64-72},
  year={2022},
  publisher={Nature Publishing Group}
}

@article{dudley2014instabilities,
  title={Instabilities, breathers and rogue waves in optics},
  author={Dudley, J. M. and Dias, F. and Erkintalo, M. and Genty, G.},
  journal={Nat. Photonics},
  volume={8},
  number={10},
  pages={755--764},
  year={2014},
  publisher={Nature Publishing Group UK London}
}

@article{berry2005tsunami,
  title={Tsunami asymptotics},
  author={Berry, Michael V},
  journal={New J. Phys.},
  volume={7},
  number={1},
  pages={129},
  year={2005},
  publisher={IOP Publishing}
}

@article{jiang2023branching,
  title={Branching of high-current relativistic electron beam in porous materials},
  author={Jiang, K and Huang, TW and Li, R and Yu, MY and Zhuo, HB and Wu, SZ and Zhou, CT and Ruan, SC},
  journal={Phys. Rev. Lett.},
  volume={130},
  number={18},
  pages={185001},
  year={2023},
  publisher={APS}
}

@article{garnier2025stochastic,
  title={Stochastic Dynamics of Incoherent Branched Flows},
  author={Garnier, Josselin and Picozzi, Antonio and Torres, Theo},
  journal={Phys. Rev. Lett.},
  volume={134},
  number={22},
  pages={223803},
  year={2025},
  publisher={APS}
}

@article{bender2022coherent,
  title={Coherent enhancement of optical remission in diffusive media},
  author={Bender, Nicholas and Goetschy, Arthur and Hsu, Chia Wei and Yilmaz, Hasan and Palacios, Pablo Jara and Yamilov, Alexey and Cao, Hui},
  journal={Proc. Natl Acad. Sci.},
  volume={119},
  number={41},
  pages={e2207089119},
  year={2022},
  publisher={National Academy of Sciences}
}

@article{xu2024modulating,
  title={Modulating the macroscopic anisotropy of liquid crystalline polymers by polarized light},
  author={Xu, Yiyi and Jin, Mengshi and Wang, Jinyu and Huang, Shuai and Li, Quan},
  journal={Responsive Mater.},
  volume={2},
  number={4},
  pages={e20240020},
  year={2024},
  publisher={Wiley Online Library}
}

@article{ma2022self,
  title={Self-assembled liquid crystal architectures for soft matter photonics},
  author={Ma, Ling-Ling and Li, Chao-Yi and Pan, Jin-Tao and Ji, Yue-E and Jiang, Chang and Zheng, Ren and Wang, Ze-Yu and Wang, Yu and Li, Bing-Xiang and Lu, Yan-Qing},
  journal={Light Sci. Appl.},
  volume={11},
  number={1},
  pages={270},
  year={2022},
  publisher={Nature Publishing Group UK London}
}

@article{wang2024moire,
  title={Moir{\'e} effect enables versatile design of topological defects in nematic liquid crystals},
  author={Wang, Xinyu and Jiang, Jinghua and Chen, Juan and Asilehan, Zhawure and Tang, Wentao and Peng, Chenhui and Zhang, Rui},
  journal={Nat. Commun.},
  volume={15},
  number={1},
  pages={1655},
  year={2024},
  publisher={Nature Publishing Group UK London}
}

@article{li2018electrically,
  title={Electrically driven three-dimensional solitary waves as director bullets in nematic liquid crystals},
  author={Li, Bing-Xiang and Borshch, Volodymyr and Xiao, Rui-Lin and Paladugu, Sathyanarayana and Turiv, Taras and Shiyanovskii, Sergij V and Lavrentovich, Oleg D},
  journal={Nat. Commun.},
  volume={9},
  number={1},
  pages={2912},
  year={2018},
  publisher={Nature Publishing Group UK London}
}

@article{zhao2025space,
  title={Space-time crystals from particle-like topological solitons},
  author={Zhao, Hanqing and Smalyukh, Ivan I},
  journal={Nat. Mater.},
  year={2025},
  doi={10.1038/s41563-025-02344-1},
  publisher={Nature Publishing Group UK London}}

@article{qincrossover,
  title={Crossover From Branched Flow to Anderson Localization in Time-Fluctuating Random Potentials},
  author={Qin, Jianwei and Liu, Yan and Ye, Fangwei},
  journal={Laser \& Photonics Reviews},
  pages={e01144},
year={2025},
publisher={Wiley Online Library}
  
}

@article{liu2025,
  title = {Observation of two-dimensional branched flow of light},
  author = {Liu, Yan and Lin, Ke and Liu, Zhaoyu and Qin, Jianwei and Fu, Qidong and Wang, Peng and Ye, Fangwei},
  journal = {Phys. Rev. Lett.},
  volume = {135},
  issue = {14},
  pages = {143801},
  numpages = {6},
  year = {2025},
  month = {Sep},
  publisher = {American Physical Society},
  doi = {10.1103/gyqw-d17z},
  url = {https://link.aps.org/doi/10.1103/gyqw-d17z}
}

@article{supplemental, 
journal={See Supplemental Material for a detailed description}}
~~
\end{document}